\documentclass[prl,aps,twocolumn ,superscriptaddress]{revtex4-1}
\usepackage{amsthm,amsfonts,graphicx,verbatim,ulem,color}
\usepackage{amssymb}
\usepackage{amsmath}
\usepackage{dsfont}
\usepackage{mathtools}
\usepackage{stackrel}
\usepackage{color}

\begin{document}

\title{Non-local Adiabatic Response of a Localized System to Local Manipulations }

\author{Vedika Khemani}
\affiliation{Department of Physics, Princeton University, Princeton, NJ 08544}
\affiliation{Max-Planck-Institut f\"{u}r Physik komplexer Systeme, 01187 Dresden, Germany}

\author{Rahul Nandkishore}
\affiliation{Princeton Center for Theoretical Science, Princeton University, Princeton, NJ 08544}

\author{S. L. Sondhi}
\affiliation{Department of Physics, Princeton University, Princeton, NJ 08544}
\affiliation{Max-Planck-Institut f\"{u}r Physik komplexer Systeme, 01187 Dresden, Germany}

\date{\today}

\begin{abstract}
We examine the response of a system localized by disorder to a time dependent local perturbation which varies smoothly with a characteristic timescale $\tau$. We find that such a perturbation induces a non-local response, involving a rearrangement of conserved quantities over a length scale $\sim \ln \tau$. This effect lies beyond linear response, is absent in undisordered insulators and highlights the remarkable subtlety of localized phases. The effect is common to both single particle and many body localized phases. Our results have implications for numerous fields, including topological quantum computation in quantum Hall systems, quantum control in disordered environments, and time dependent localized systems. For example, they indicate that attempts to braid quasiparticles in quantum Hall systems or Majorana nanowires will surely fail if the manipulations are performed asymptotically slowly, and thus using such platforms for topological quantum computation will require considerable engineering. They also
establish that disorder localized insulators suffer from a statistical orthogonality catastrophe.
\end{abstract}

\maketitle

The study of localization in isolated disordered systems has a rich history dating back to the seminal work of P. W. Anderson \cite{Anderson}. While it is well known that strong enough disorder exponentially localizes single-particle wavefunctions, the fate of interacting many-body systems in disordered landscapes remains a long-standing problem \cite{Fleishman}. Recent progress \cite{BAA, agkl, Mirlin}  on the phenomenon of many-body localization (MBL) has led to an intense revival of interest in this subject \cite{imbrie, Oganesyan, pal, Znid} -- for a review, see Ref. \onlinecite{arcmp}. MBL phases have a rich complex of properties including (i)  vanishing long wavelength conductivities at finite temperatures \cite{BAA}, (ii) an extensive number of local conserved quantities \cite{Lbits, Abanin}  leading to a breakdown of ergodicity 
and  (iii)  spectral functions of local operators that show a `mobility' gap   at all temperatures \cite{ngh, johri}. Strikingly,  (iv) highly excited MBL eigenstates can exhibit localization protected order -- both Landau symmetry-breaking and topological order --  in dimensions and at energy densities normally forbidden by the Peierls-Mermin-Wagner theorem \cite{LPQO, Pekker, Vosk, Kjall, Bauer, Bahri, lspt, qhmbl}. 

MBL systems present the tantalizing possibility of using localization to protect quantum computation. Localized systems might serve as protected quantum memories since they undergo only slow (logarithmic) dephasing, and even this can be removed by spin echo procedures \cite{Bauer, DEER, Sid, ngh}.  Prima facie, one expects to be able to locally manipulate degrees of freedom in such systems without affecting distant q-bits,  a property with various quantum control applications.  Further, property (iv) above raises the interesting possibility of performing topological quantum computation at finite temperatures in the MBL regime by braiding excitations in topologically ordered MBL eigenstates. While there isn't an energy gap at finite temperatures, the `mobility gap' could serve to protect adiabatic braiding instead.

\begin{figure}[htbp]
\includegraphics[width=\columnwidth]{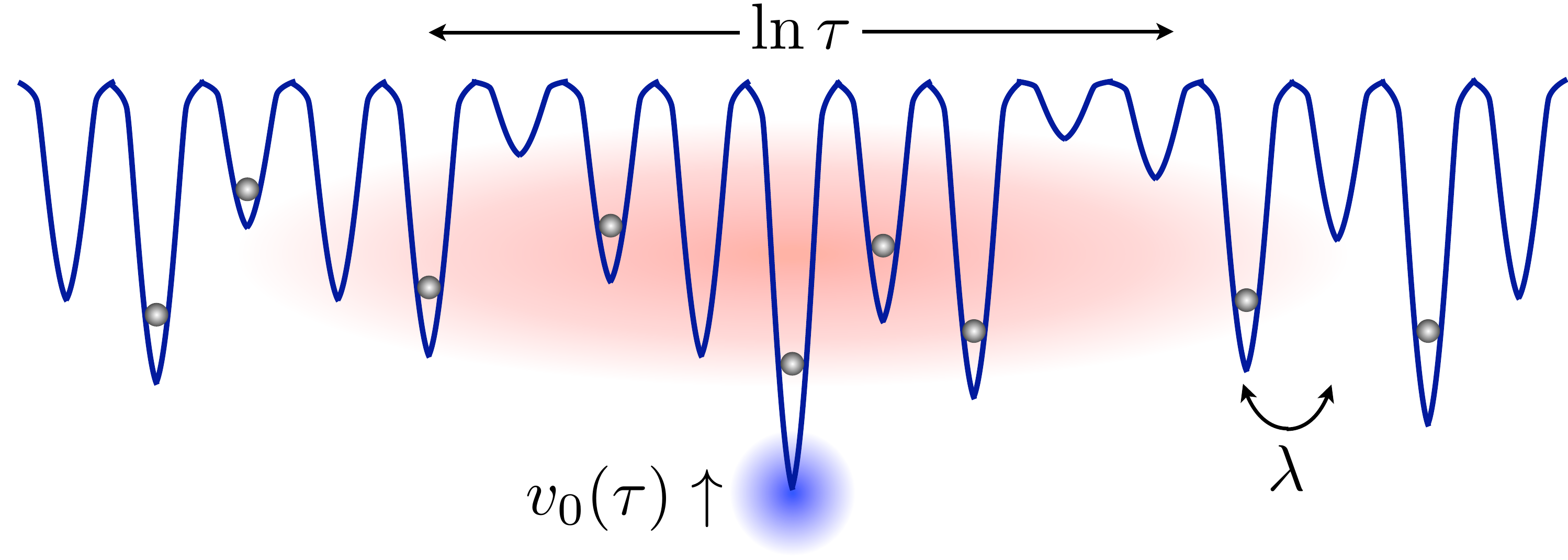}
\caption{Schematic illustration of our protocol. A \textit{local} time-dependent potential $v_0(\tau)$ leads to a highly non-local adiabatic charge response in disordered insulators, causing a `zone of disturbance' with radius $\sim \ln \tau$.  }
\label{Fig:Schematic}
\end{figure}

\begin{figure*}[htbp]
\includegraphics[width=\textwidth]{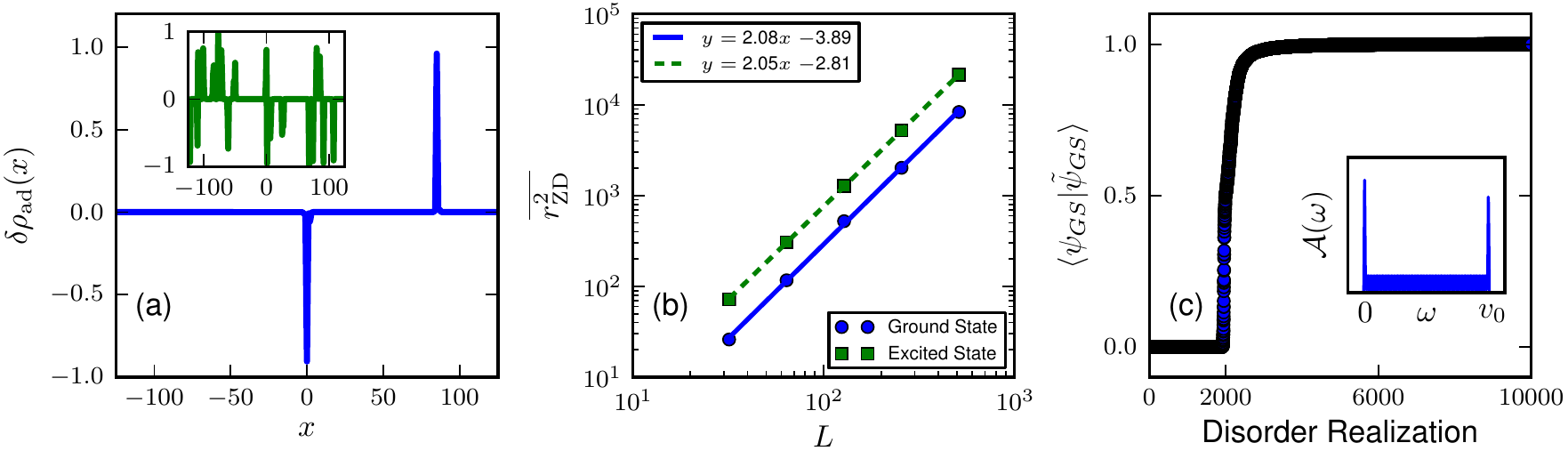}
\caption{ (a) Adiabatic change in the ground state charge density (Eq.~\eqref{Eq:DeltaRho}) in a given disorder realization in a 250 site Anderson model (Eq.~\eqref{Eq:FFHopping}) subject to a repulsive time-dependent potential on site 0 with $\lambda/W = .1$. Charge is expelled from site 0 and transferred to a distant location (near site 75). Inset: Same for an excited state in the middle of the spectrum (colloquially, a $T = \infty$ excited state) showing a multi-particle rearrangement over a large `zone of disturbance'. (b) Scaling of the radius of zone of disturbance (Eq.~\eqref{Eq:RZD}) with system size for the ground state (blue circles) and $T= \infty$ excited states (green squares) averaged over $10^4$ disorder realizations showing a linear scaling $r_{ZD} \sim L$ in both cases.   (c) Overlaps of the MB ground states in the presence ($|\tilde{\psi}_{GS}\rangle$) and absence ($|\psi_{GS}\rangle$) of a local potential of strength $v_0 = .4$ at site 0 sorted over  $10^4$ disorder realizations with $f = v_0/2W = .2$. The sorted values show a \textit{statistical} orthogonality catastrophe with probability $f = .2$. Inset: Sketch of the disorder averaged spectral function (Eq.~\eqref{Eq:Xray}) for Anderson insulators. The spectrum is pure point, and the diffuse, non-zero strength between $(0, v_0)$ is a signature of the long-distance charge rearrangement.
}
\label{Fig:ChargeTransfer}
\end{figure*}

All these applications require local manipulation of quantum degrees of freedom. Motivated by these considerations, we study the adiabatic response of localized phases to local perturbations using a combination of analytic arguments and numerical exact diagonalization. In particular,  we study  the adiabatic time evolution of a system governed by the time dependent Hamiltonian $$H(t) =  H_{\rm L} + V(t/\tau),$$ where $H_{\rm L}$ is a  localized Hamiltonian, $V$ is a time dependent local perturbation which acts only on a small compact subregion in real space, and which is zero in the distant past and future ($ t\rightarrow \pm \infty$). Finally,  $\tau$  sets the time scale on which the perturbation changes. In this work, adiatabatic time evolution will be understood to mean
$$
|\psi(t)\rangle =  \lim_{ L \to \infty} \lim_{\tau \to \infty} U(t/\tau) |\psi(0)\rangle
$$
where $U(t)$ is the unitary time-evolution operator defined by $H(t)$, and the limit $\tau \to \infty$ is taken before the thermodynamic limit $L \to \infty$. We also discuss the opposite order of limits.

A naive understanding of localization suggests that the influence of the perturbation $V(t)$ should be spatially confined to within a localization  length $\xi$ of the region in space where $V$ acts. The recent discovery \cite{Bardarson}  of logarithmic dephasing and entanglement spreading in interacting, localized systems 
updates this understanding, but nonetheless leaves in place the intuition that conserved charges, such as number and energy,
should not move over distances greater than the localization length. We will show that this understanding needs to be further updated. 

Our main results are as follows: (a) A local perturbation remarkably induces a highly non-local adiabatic {\it charge} response in distant parts of the system.  For an infinitely slow perturbation, $\tau \rightarrow \infty$, there is a ``zone of disturbance'' where charge rearrangement occurs over a length scale that diverges linearly with system size. For finite $\tau$, charge transfer takes place over length scales $\sim \log (\tau)$.  See Fig. \ref{Fig:Schematic}. This effect is distinct from the logarithmic entanglement growth as the charge spreading occurs even in the non-interacting problem where there is no entanglement spreading. (b) This effect cannot be captured by linear response theory, and revises our understanding of susceptibility and transport in localized phases. Our results also modify our understanding of MBL in time dependent systems \cite{Luca, Ponte}. And (c), there is a \textit{statistical} Anderson orthogonality catastrophe \cite{AndersonOC}  for both ground and highly-excited states in strongly localized systems, contrary to established wisdom for ground states \cite{borisOC}. 
Importantly, our work places strong constraints on possibilities for quantum control and topological quantum computation in disordered systems as we will discuss below. We note that there are parallels to our discussion of local manipulations of disordered systems in the field of optics \cite{optics1, optics2}.

\textbf{Anderson Insulator --} 
We start with a disordered single-particle (SP) Anderson insulator in 1D with a time-dependent local potential where most of our results can be described in a transparent setting. Generalizations to higher dimensions is straighforward. Many-body (MB)  eigenstates are constructed by simply filling the SP levels. To characterize the non-local charge response of MB eigenstates, define the adiabatic change in the charge density as: 

\begin{align}
\delta \rho_{\rm ad} (x) = \sum_{\alpha \; \rm occ} |\psi_{\alpha} (x, t = \infty)|^2 - |\psi_{\alpha} (x, t = -\infty)|^2 
\label{Eq:DeltaRho}
\end{align}
where $\psi_{\alpha}(x,t)$ is the $\alpha$-lowest instantaneous SP eigenstate of $H(t)$, and the sum is over occupied SP states in a given MB eigenstate. Figure~\ref{Fig:ChargeTransfer}(a) shows $\delta \rho^{\rm ad}$ for the MB ground state and an excited state (drawn randomly from the infinite temperature Gibbs ensemble) in a given disorder realization; both show a long-distance rearrangement of charge. We emphasize that this transfer is mediated by the action of a strictly local potential in an insulator!
More precisely, define
\begin{align}
r_{\rm ZD}^2 =  \frac{\int_{-L/2}^{L/2} dx \; x^2 \; \delta \rho_{\rm ad}^2}{\int_{-L/2}^{L/2} dx \;  \delta \rho_{\rm ad}^2}
\label{Eq:RZD}
\end{align}
where $L$ is the system size and $r_{\rm ZD}$ quantifies the radius of the zone of disturbance over which charge rearrangement takes place. It would be natural to expect $r_{\rm ZD}$ to scale as the localization length $\xi$. Instead, we find that the disorder averaged radius diverges linearly with system size,   $\sqrt{\overline{r_{\rm ZD}^2}} \sim L $, \textit{i.e.} the zone of disturbance grows without bound in the adiabatic limit.  Fig~\ref{Fig:ChargeTransfer}(b) shows the disorder averaged scaling of $\overline{r_{\rm ZD}^2}$ with system size for both the MB ground state and $T = \infty$ excited states. 

\begin{figure}[htbp]
\includegraphics[width=6.5cm]{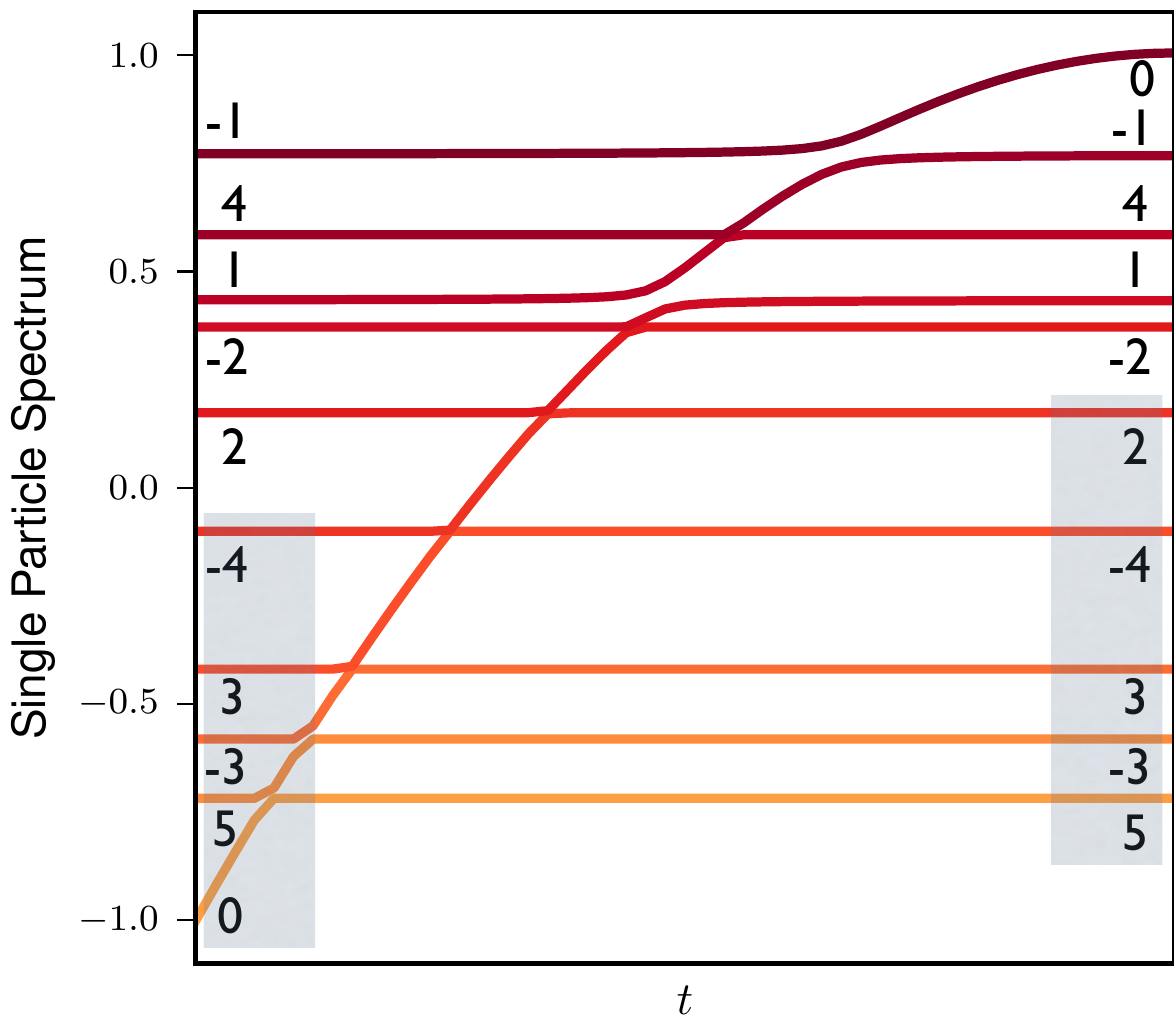}
\caption{Single-particle spectrum of a 10 site Anderson insulator, Eq.~\eqref{Eq:FFHopping}, in a given disorder realization as a function of time. The numbers denote the localization centers of the corresponding eigenstates. The changing potential on site 0 brings $|\psi\rangle_0(t)$ into resonance with the other eigenstates and leads to a set of avoided crossings. The MB ground states at half filling (shaded levels occupied) in the distant past and future are related through the transfer of charge from site 0 to site $R_m = 2$.}
\label{Fig:LevelCrossing}
\end{figure}

To understand these results, let's turn to the specific fermionic Hamiltonian in which our computations were performed: 
\begin{align}
H(t) & = H_{\rm hopping} + V_{\rm loc}(t/\tau) \nonumber \\ 
& =   \sum_{i= -L/2}^{L/2-1} -\lambda ( c^{\dagger}_i c_{i+1} + c^{\dagger}_{i+1}c_i) + v_i c^{\dagger}_i c_i + v_0(t/\tau) c^{\dagger}_0 c_0
\label{Eq:FFHopping}
\end{align}
where $\lambda$ is the nearest-neighbor hopping strength, the onsite potentials $v_i$ are drawn uniformly from $[-W, W]$ and $v_0(t)$, the potential on site 0 is changed with time. We now focus on the strong disorder limit, $\lambda/W \ll 1$, where the localization length $\xi \sim 1$ is on the scale of a lattice constant and an especially simple picture emerges. Denote the eigenstate with localization center at site  $r$ as $|\psi\rangle_{r}$. As $v_0(t)$  is varied in time, the eigenenergy of  $|\psi\rangle_{0}$ is affected most strongly. To leading order, as $v_0(t)$  sweeps the range from $-W $ to $W$, $|\psi\rangle_{0}(t)$ comes into resonance with each of the other eigenstates giving rise to a set of avoided crossings with gaps that scale as $~ \lambda \exp\big(-R \ln(W/\lambda)\big)$, the effective coupling between $|\psi\rangle_{0}$ and $|\psi\rangle_{\pm R}$. Thus, the smallest gaps (due to a resonance between $|\psi\rangle_{0}$ and $|\psi\rangle_{O(N)}$) scale exponentially with system size ($\sim \lambda^N$) even though the system is non-interacting, a fact previously discussed \cite{Altshuler} in the context of adiabatic quantum optimization (though for a global perturbation). 
Figure~\ref{Fig:LevelCrossing} shows the evolution of the spectrum for a given disorder realization in a 10 site chain with $W = 1$.  We note that such resonances are also present in an unperturbed Anderson insulator with exponentially small probability; however, the local drive ensures that they occur with probability one.

The many-body ground state of fixed number, say $m = N/2$, is constructed by filling the $m$ lowest single-particle states Thus, if the system evolves adiabatically, a purely local perturbation on site 0 leads to a transfer of charge a distance $R_m$ away, where $|\psi\rangle_{R_{m}}$ is the $m^{th}$ lowest eigenstate in the distant future! See Fig.~\ref{Fig:LevelCrossing}. The value of $R_m$ differs between disorder realizations, but can take any value from $1$ to $N/2$ with equal probability in a system with uniform disorder strength. Thus, the disorder averaged response to the local perturbation has a very wide spatial distribution, and shows no decay on scales longer than $\xi_{\rm loc}$. When $v_0$ sweeps only a finite fraction $f$ of the bandwidth ($\sim W$), distant charge transfer in the ground state happens with probability $f$, occurring only if an occupied state is swept through an avoided crossing with an unoccupied state. The disorder averaged response still shows no decay. For highly excited MB states, the adiabatic response leads to a multi-particle charge rearrangement in a diverging zone of disturbance as shown in Fig.\ref{Fig:ChargeTransfer}(b).

Having characterized the spatial spread of the adiabatic response, we now turn to the ramp time $\tau$ needed for adiabatic time evolution. In particular, we want to know whether $\tau$ is set by the exponentially small avoided crossing gaps or by an $O(1)$ mobility gap (\cite{ngh}). 

For the $n^{th}$ SP eigenstate of $H(-\infty)$ to remain the $n^{th}$ instantaneous eigenstate of $H(t)$, the adiabaticity condition
\begin{align}
a_{mn} (t)= \hbar \;\frac{  \langle \psi_m(t)|\frac{\partial V(t/\tau)}{\partial t} |\psi_n(t) \rangle}{(E_m(t) - E_n(t))^2}  \ll 1
\end{align}
must be satisfied at all times for all $m \neq n$, where the eigenstates are defined by $H(t) \psi_\alpha(t) = E_\alpha(t) \psi_\alpha(t)$. 
For a local $V(t/\tau)$, one might expect the numerator of $a_{mn}$ to be significant only when $\psi_{m,n}$ are centered within a few localization lengths of each other and the potential; 
however, states within a localization volume in space are separated in energy by the mobility gap giving a large denominator. Thus, naively $a_{mn} \ll 1$ so long as $\hbar/\tau$ is smaller than the mobility gap.

This reasoning fails at the avoided crossings in our locally perturbed system. At an avoided crossing at time $t$ between eigenstates $|\psi (t)\rangle_{0}  $ and $|\psi(t)\rangle_{R}$ the energy gap is exponentially small in $R$, wheras the instantaneous eigenstates look like the symmetric and anstisymmetric combinations : $|\psi_{m,n}(t)\rangle \sim |\psi(t)\rangle_{0} \pm|\psi(t)\rangle_{R}(t)$. Since $V(t)$ is also localized near site $0$, the numerator of $a_{mn}(t)$ receives a substantial contribution from the diagonal piece ${}_0\langle \psi(t) | \dot{V} |\psi(t)\rangle_0$.  Thus, the system remains adiabatic only if 
$$ \tau_{\rm ad} \gg \frac{\hbar W^{2(R-1)} \partial_t V }{\lambda^{2R}}$$
i.e. the mobility gap does not protect adiabaticity.  Thus in a system of size $L$, the drive is adiabatic for all levels only if the ramp is exponentially slow in the system size, even for a single-particle Anderson insulator. 

The preceding discussion also implies that with a finite ramp time $\tau$, the system is only able to adiabatically avoid level crossings with gaps $> \hbar/\tau$. Since the charge transfer is a consequence of avoided crossings and since the gaps decay exponentially with distance $W \exp(-R/\xi)$ (in the strong localization regime),  with a finite ramp time $\tau$, charge transfer occurs over a characteristic length scale 
\begin{equation}
{r}_{\rm ZD} \sim \xi \ln \left(\frac{\tau W}{\hbar} \right) \sim \ln(\tau)
\end{equation}
This logarithmic transfer of charge is our key result. We predict a similar logarithmic spreading in the weak localization regime, on distances larger than the localization length. 

\textit{Orthogonality Catastrophe --} 
This non-local charge response implies a \textit{statistical} Anderson orthogonality catastrophe (OC) in the Anderson insulator. Anderson's original work had shown that the many-body ground states of a clean (metallic) system of fermions in the presence and absence of a local impurity potential were orthogonal in the thermodynamic limit, even for arbitrarily weak (but finite) potentials. In the strongly disordered system under study, adding an on-site potential on site 0 of strength $v_0 = f*(2W)$ with $f < 1$ leads to a distant charge transfer and hence an orthogonal new ground state with probability $f$. Figure~\ref{Fig:ChargeTransfer}(c) shows the ground state overlaps with and without a potential, clearly showing an orthogonality with probability $f$ (roughly when the starting potential on site 0 lies within $v_0$ of the Fermi energy). For highly excited states, we have a catastrophe with probability 1. Previous work \cite{borisOC} on the OC in ground states of strongly disordered sytems only captured the non-orthogonal overlaps that occur with probability $1-f$ to incorrectly conclude that strongly disordered insulators don't suffer from the OC.

The OC has important consequences for several dynamical phenomena in metals. Famously, it predicts an X-ray edge singularity \cite{Hopfield, vonDelft}, where the low-energy X-ray absorption spectrum in a metal has the singular form $\mathcal{A}(\omega) \sim w^{-1 + 2\eta}$ and $\eta$ is derived from the Anderson OC. The primary spectral function characterizing local quantum quenches (such as a change in the potential) takes the form: 
\begin{align}
\mathcal{A}(\omega) = \sum_n |\langle n | GS \rangle |^2 \delta(\omega - E_n + E_{GS} + \omega_0)
\label{Eq:Xray}
\end{align}
 where $|GS\rangle$ is ground state of the system before the quench, and $|n\rangle$, $E_n$ are the eigenstates and eigenvalues of the final Hamiltonian. For the Anderson insulator in the $ \lambda/W \ll 1$ limit, $A(\omega)$ looks pure point, and is characterized by delta function peaks located between $\omega \in (0, v_0)$ for different disorder realizations with a catastrophe. In disorder realizations where there is no catastrophe and no long-distance charge transfer (probability $1-f$), $\mathcal{A}(\omega)$ has a peak at either $\omega = 0$ or $ \omega = v_0$. The inset in Fig.~\ref{Fig:ChargeTransfer}(c) shows a representative sketch of the disorder averaged spectral function -- the non-zero weight between $(0, v_0)$ distinguishes the Anderson insulator response from that of ordinary band insulators and is a signature of long-distance charge rearrangement. 

\begin{figure}[htbp]
\includegraphics[width=\columnwidth]{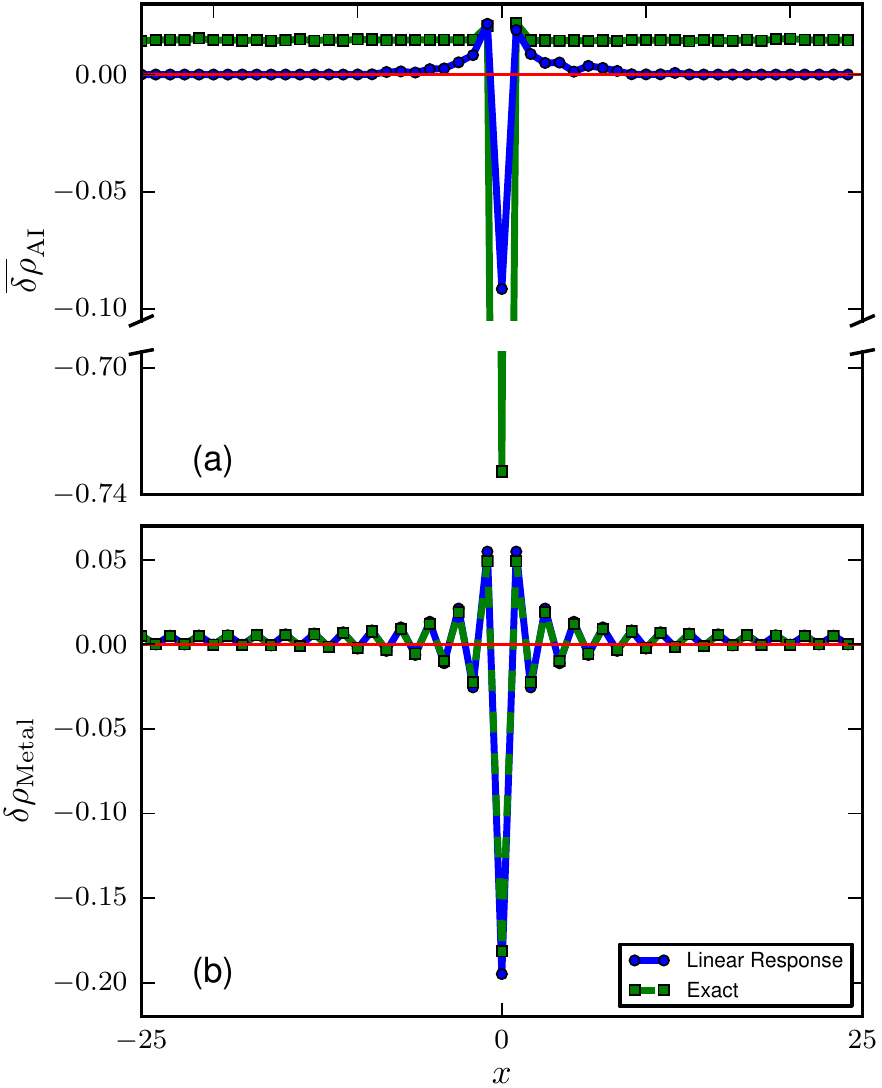}
\
\caption{ Exact (green, squares) and linear-response (blue, circle) answers for the ground-state charge density difference (Eq. \eqref{Eq:DeltaRho}), ${\delta\rho} (r)$, in a system of size $L = 50$    due a perturbing repulsive potential of strength $v_0 = .4$  added to the center of the system in (a) an Anderson insulator with $\lambda/W = .1$ averaged over $10^5$ disorder realizations. While $\delta\rho_{\rm LR} (r)$ rapidly decays away from the location of the potential, the exact  $\delta\rho_ (r)$ shows a uniform response everywhere (with amplitude scaling as $1/L$). (b) In a metal ($\lambda = 1, W = 0$), the linear-response charge response closely captures the exact answer. 
}
\label{Fig:LinearResponse}
\end{figure}

\textit{Failure of linear response --} Before leaving the Anderson insulator it is instructive to compare our description of the adiabatic response to a local perturbation to the more standard account of such a perturbation in linear response (LR) theory in the $\omega \rightarrow 0$ limit.  This response is governed by the density susceptibility which has been calculated using LR, for example, in the classic work by Vollhardt and W\"olfle \cite{Vollhardt} and the LR answer {\it is} local on the scale of the localization length. As the standard computations are approximations carried out with disorder averaged Green's functions and done at fixed chemical potential, it is useful to revisit this question more carefully. For a single disorder realization the density susceptibility is given by a Kubo formula which involves matrix elements of the perturbation between the exact {\it unperturbed} eigenstates. However, the unperturbed eigenstates will {\it not} be part of a long range resonance (with probability exponentially near 1) wheras the long range transfer of charge occurs {\it only} because the {\it exact} (perturbed) eigenstates are tuned through a long range resonance. Thus the linear response result will indeed be local, in contrast to the adiabatic result. This has been verified by numerical computations as shown in  Fig.~\ref{Fig:LinearResponse}(a). Finally, readers concerned that our diverging length scale at large $\tau$ is somehow 
related to Mott's celebrated formula  $\sigma \sim \omega^2 \ln^{d+1} \omega$ \cite{Mott} for the AC conductivity at small $\omega$ should note that this is now ruled out as the formula is basically an excercise in linear response
theory.

\textbf{Contrast with Clean Insulators and with Metals -- }
At this point it instructive to contrast the behavior we have found for the Anderson insulator with that of undisordered insulators (band and Mott) and that of metals. In an undisordered insulator the particles are again localized with a length scale that can be read off from correlation functions and which will scale inversely with the gap. Now a) the response to an adiabatically prepared local potential is indeed localized with this localization length, b) there is no orthogonality catastrophe, c) the adiabatic response is accurately captured by linear response/perturbation theory.
The case of metals---ballistic and diffusive---is intermediate. In a metal a) charge can flow to infinity and thus the adiabatic charge transfer is not restricted to the vicinity of the applied perturbation, b) there {\it is}---famously---an orthogonality catastrophe with a scaling with system size that is modified in the diffusive case, c) the adiabatic response is accurately captured by linear response/perturbation theory as illustrated in~\ref{Fig:LinearResponse}(b).
As a function of the time scale we can be more specific.  In both ballistic and diffusive metals we will
obtain a power law spreading of charge $R_\tau \sim \tau^{\sigma}$ with $\sigma=1,1/2$ respectively. Indeed, charge can 
continue to flow even long after the Hamiltonian stops changing ($t \gg \tau)$ allowing the effects of a local perturbation to propagate out to infinity. This is in contrast to both  undisordered insulators and Anderson localized systems, where charge transfer occurs {\it only} when the Hamiltonian is changing, and thus the influence of the perturbation is restricted to a finite region of space (with linear size $\tau^0$ or $\ln \tau$ respectively). Moreover, the smallest gaps in non-interacting metals scale only polynomially with system and thus an adiabatic response can be achieved by much faster ramp rates $\tau$ as compared to localized systems. Finally it is interesting to note that in clean insulators and metals the adiabatic limit considered in this paper yields the same charge response
as the opposite limit in which $L  \rightarrow \infty$ before $\tau \rightarrow \infty$. In the Anderson insulator the latter limit 
fails to exist as a consequence of the physics discussed here.

\textbf{Generalization to MBL--} We now generalize our analysis to fully MBL interacting localized systems. Our principal results
for the Anderson insulator carry over. MBL systems also exhibit a) a zone of disturbance that grows as $\ln(\tau)$, b) a statistical orthgonality catastrophe for the ground state and a certain orthogonality catastrophe for highly excited states and c) a failure of linear reponse to agree with this behavior. There are three new features that come into play. First, we can consider systems that lack a parent single particle description as they lack a conserved number and the non-local response now involves a rearrangment in the energy density alone. Second, the rearrangement process for highly excited states now exhibits a range of length scales from the ubiquitous $\ln(\tau)$ to the shorter, but still divergent, $(\ln \tau)^{1/(d+1)}$ at which much more comprehensive changes take
place in the structure of the ground state. Third, the termination of the perturbation is now followed by the entanglement spreading discussed in \cite{Bardarson}.  We note that the entanglement spreading is the dominant effect in the recent work on quantum revivals \cite{Sid} which considers sudden quenches and thus works in the opposite limit from the one considered here.  
These results can be derived within the ``l-bits'' formalism introduced in \cite{Abanin, Lbits} and the reader is referred to the Supplementary Material for more details. 
For specificity consider a quantum 
spin system dominated by random fields:
\begin{equation}
H = \sum_{i=-L/2}^{L/2} h_i \sigma^z_i + \lambda H^{(2)} 
\end{equation}
where the $h_i$ are taken from a distribution of width $W$, $H^{(2)}$ includes interaction terms that may or not conserve
$\sum_i \sigma^z_i$ and $\lambda$ is chosen appropriately small so that the eigenstates of $H$ are localized at {\it all} 
energies. Such a fully MBL system can be recast via a unitary transformation into the form \cite{Abanin, Lbits}
\begin{equation}
H_0  =  \sum_i \tilde h_i\tau^z_i + \sum_{i,j}\tilde J_{ij}\tau^z_i \tau^z_j
 + \sum_n \sum_{i,j,\{k\}}K^{(n)}_{i\{k\} j}\tau^z_i\tau^z_{k_1}...\tau^z_{k_n} \tau^z_j ~.
 \label{eq:H_u}
\end{equation}
where the $\tau^z_i$ are exponentially localized emergent integrals of motion (`l-bits'), and the high order terms $J_{ij}$ and $K_{i\{k\}j}$ fall off exponentially rapidly in the range, modulo exponentially rare resonant ones which can be ignored for
the most part. To leading order in large $W$, the $\tau^z_i$  coincide with the $\sigma^z_i$, but with a `dressing' of multi-spin operators that falls off exponentially in the range.

Let us now introduce a local perturbation by making the field on a particular site $h_0$ time dependent $h_0 \rightarrow h_0(t/\tau)$.  With this change the new $\tilde h_0$ and the interaction terms involving $\tau_0$ also become time dependent. Further, the other l-bits $\tau_{i\neq0}$ will also be affected due to their overlap with $\sigma_0$. In particular, the l-bits themselves will have to be redefined continuously in time, so that (written in terms of the l-bit operators $\tau^z_i$ at time zero), the Hamiltonian will acquire off diagonal terms:
\begin{eqnarray}
H(t>0) &=& H_0 +  \sum_i \tilde h_i \exp(-|i|/\xi_0) \tau^z_i + \sum_{i}\tilde J_{ij}(t) \tau^z_i \tau^z_0 \nonumber\\
 &+& \sum_{n, i,\{k\}}K^{(n)}_{i\{k\} j} (t) \tau^z_i\tau^z_{k_1}...\tau^z_{k_n} \tau^z_0 \nonumber\\
&+&  \sum_{j} (t^x_{j0}(t) \tau^x_j \tau^x_0 + t^y_{j0}(t) \tau^y_j \tau^y_0 ) + ...
\end{eqnarray}
where the $...$ denotes higher order l-bit spin hopping terms which rearrange multiple l-bits and the off diagonal terms all fall off exponentially with distance from $0$, both in the magnitude of individual terms and in the total weight.

Now, a slowly varying time dependent potential induces avoided crossings, with the minimum gaps controlled by the off diagonal terms. Range $R$ single l-bit spin hops will then occur IFF the Hamiltonian varies slowly compared to the gap scale $\exp(-R/\xi_0)$, where $\xi_0$ is the characteristic length scale for decay of a typical off diagonal term. The effect of the higher order interaction terms is to modify the effective interaction length so that the gaps fall off as $\exp(-R/\tilde\xi)$, where $\tilde \xi$ is the decay length of the total interaction. The largest gaps are typically set by {\it many spin} rearrangements. Thus, the size of the zone of disturbance (the region over which some l-bits are rearranged) grows as $\tilde \xi \ln \tau$, with $\tilde \xi \ge \xi_0$. For highly excited states we can also identify a ``zone of total rearrangement'' - a (smaller) region of size $\tilde R$ over which a $\tilde R$ independent fraction of the l-bits are rearranged. Since the number of l-bits that must be rearranged in the zone of total rearrangement grows as $\tilde R^{d}$, and the matrix element for flipping each l-bit is exponentially small in $\tilde R$, the gaps associated with total rearrangements will scale as $\exp(-\tilde R^{d+1})$. Thus, we expect the zone of total rearrangement to grow as $\tilde R \sim (\ln \tau)^{1/(d+1)}$. One can also establish the remaining results within the same framework.

\begin{figure}[htbp]
\includegraphics[width=\columnwidth]{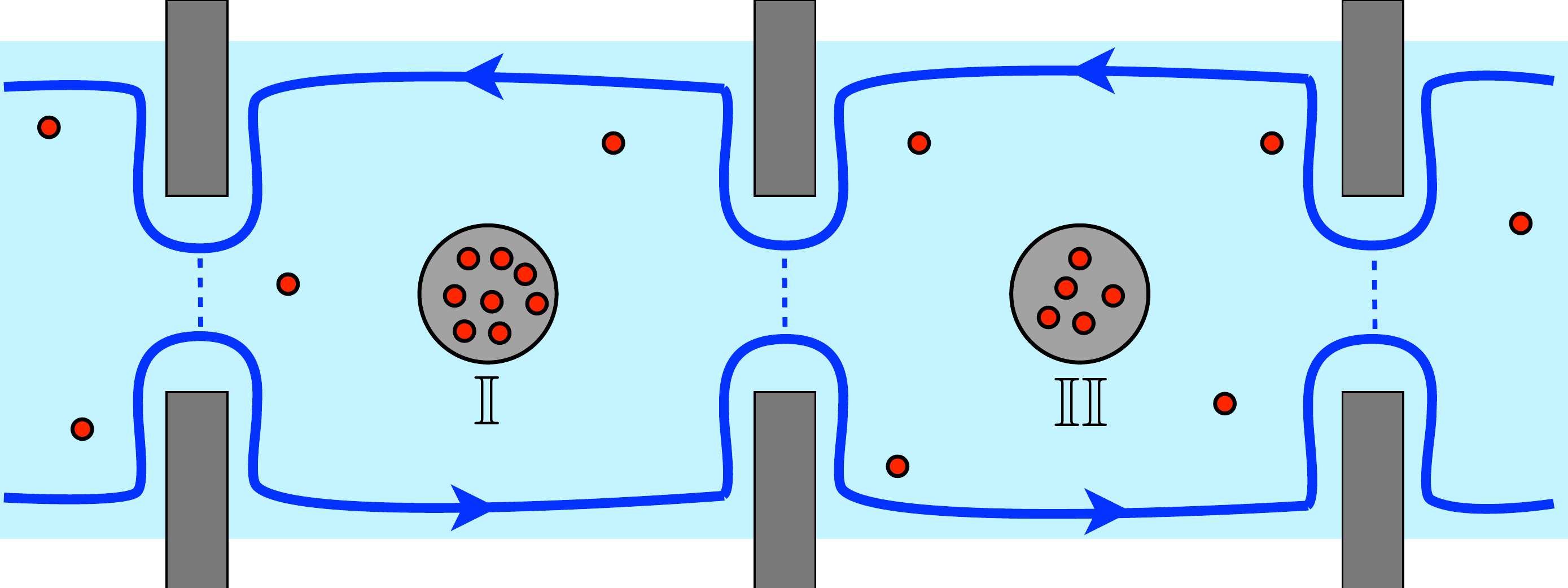}
\
\caption{ Schematic illustration of a proposal for topological quantum computation outlined in Ref. \onlinecite{NayakRMP}. Regions $\mathds{I}$ and $\mathds{II}$ contain the intended non-Abelian quasiparticles whose joint state is measured by interferometric tunneling experiments of quasiparticles across the constrictions. In realistic experiments, there will be unintended quasiparticles in the shaded region outside of $\mathds{I}$ and $\mathds{II}$ which will rearrange over long distances in response to the changing potential on the constriction, thereby spoiling the braiding experiment.  
}
\label{Fig:QuantumComputation}
\end{figure}

\textbf{Discussion and Ramifications--} 
We conclude by discussing the implications of our work for experiments, other aspects of the physics of localized systems and for quantum engineering. Starting with experiments, it would be gratifying to directly observe the zone of disturbance created in response to a local perturbation and it seems to us that the cold atomic systems which have exhibited Anderson \cite{AspectPT, AspectNature, DeMarcoScience, IgnuscioNature} and, apparently, MBL \cite{DeMarcoMBL} are the best places to look. 

Alternatively experiments could look for the predicted form of the X-ray absorption spectrum sketched in Fig.~\ref{Fig:ChargeTransfer}(c).  Apart from solids hosting disordered electron gases \cite{Shahar}, cold atomic gases \cite{DemlerOC} are again plausible sytems to observe this effect. 

The tuned resonance behavior that underlies our chief results can, in principle, be produced in other ways---e.g. by sandwiching a localized system between two conducting leads and tuning the chemical potential in the conducting regions. Indeed, this is the well known setting of the Lifshitz-Azbel \cite{Azbel, Pendry, Lifshitz} resonances in Anderson localized systems and a natural extension is to look for generalizations of these to MBL systems (work in progress). Quasiperiodic systems, with \cite{HuseQP} and without \cite{AubryAndre} interactions, are known to exhibit localized states and are natural for studying our results in a setting without disorder (work in progress).

This resonance behavior has one negative implication for MBL physics though. As outlined in \onlinecite{LPQO},
MBL eigenstates can exhibit topological order and one might wonder whether it is possible to diagnose such order by means
of quasiparticle braiding as in the clean limit. Our results here imply that one {\it cannot} define an adiabatic process with a well
defined Berry phase and so quasiparticle statistics are, as such, ill-defined in the localized setting (which includes the 
ground states that underlie quantum Hall plateaux). Instead the topological information of the parent topological states must be reconstructed from other data as we will discuss elsewhere (work in progress). Another problem for which our results have consequences is that of Floquet localization in MBL systems. It has been argued \cite{Luca, Ponte} that MBL systems subject to a periodic local driving do not absorb energy indefinitely. In particular, the eigenstates of the Floquet operator for such systems are expected to be MBL. Here our results predict that slow, low-frequency local drives (or a slow perturbation of the amplitude of a fast drive) will give rise to a diverging `zone of disturbance' in the Floquet eigenstates and lead to a transfer of energy deep into the system.

Finally we turn to the implications of our work for quantum control, engineering and computation where it might often be neccessary,
for practical reasons, to seek to perform local manipulations in disordered environments while leaving distant regions untouched. At the broadest level our results imply that such control will be problematic if we attempt to carry out such manipulations arbitrarily gently/slowly as one might wish to for a theoretical analysis of devices. We emphasize that such adiabaticity is implicit in thinking of
ideal control by means of gates, for example, or even of small excitation currents which imply slow changes of various potentials.
For concreteness let us comment on a proposal to use quantum Hall systems as platforms for topological quantum computation
by creating and braiding localized excitations. In the very simplest setting shown in Fig.~\ref{Fig:QuantumComputation}, taken from Ref. \onlinecite{NayakRMP}, a qubit is created from two localized non-Abelian quasiparticles localized in regions $\mathds{I}$ and $\mathds{II}$ whose joint
boundary is defined by two constrictions used interferometrically to detect the joint state of the particles and hence of the
qubit. For our purposes it is enough to focus on the third constriction---which separates $\mathds{I}$ and $\mathds{II}$---which
is turned on and off in order to tunnel a quasiparticle between the edges and thus flip the qubit. This constriction will arise 
from an electrostatic potential with a dipolar shadow leaking into regions $\mathds{I}$ and $\mathds{II}$. In a completely
ideal device with no localized quasiparticles apart from the ones created by the experimenter, this time-dependent potential
in regions $\mathds{I}$ and $\mathds{II}$ will have no effect as long as it is not too big in magnitude. However for most
realistic devices, and {\it all} the ones that exhibit a quantum Hall plateaux prior to patterning, there will be additional localized
quasiparticles which will be subjected to this potential and can then rearrange in response and cause the computational step to fail {\it especially} if the gate is operated arbitrarily slowly. Thus, there is no safe asymptotic limit, and braiding in these devices will require considerable engineering. Similar problems will bedevil attempts to perform topological quantum computation in Majorana nanowire networks \cite{Sau, Alicea}, in the presence of localized gapless Majorana excitations and quasiparticles, which will likely also be present in realistic samples. 

Altogether, our results place natural limits on the manipulation of local degrees of freedom in localized phases and help further elucidate the remarkably subtle nature of localization.

 {\it Acknowledgements}: We acknowledge useful conversations with  B. L. Altshuler, P. W. Anderson,  J. E. Avron,  Ravin Bhatt, A. Elgart, M. S. Rudner and, especially, John Chalker. We thank John Chalker and David Huse for comments on a draft. This
work was supported by NSF Grant Numbers DMR 1006608, 1311781 and PHY-1005429 and the John Templeton Foundation (VK
and SLS) and by a PCTS fellowship (RN).

\end{document}